\documentclass[aps,prd,twocolumn,groupedaddress]{revtex4-2}
\usepackage{graphicx}
\usepackage{hyperref} %hyperlink your equations and figures and sorts

\begin{document}
%\linenumbers
%\preprint{}
\title{Can we constrain the extragalactic magnetic field\\ from very high energy observations of GRB~190114C?}

\author{T.A. Dzhatdoev}
\email[]{timur1606@gmail.com}
%\thanks{}
\affiliation{Federal State Budget Educational Institution of Higher Education, M.V. Lomonosov Moscow State University, Skobeltsyn Institute of Nuclear Physics (SINP MSU), 1(2), Leninskie gory, GSP-1, 119991 Moscow, Russia}
\affiliation{Institute for Cosmic Ray Research, University of Tokyo, 5-1-5 Kashiwanoha, Kashiwa, Japan}

\author{E.I. Podlesnyi}
\email[]{podlesnyi.ei14@physics.msu.ru}
%\thanks{}
\affiliation{Federal State Budget Educational Institution of Higher Education, M.V. Lomonosov Moscow State University, Department of Physics, 1(2), Leninskie gory, GSP-1, 119991 Moscow, Russia}
\affiliation{Federal State Budget Educational Institution of Higher Education, M.V. Lomonosov Moscow State University, Skobeltsyn Institute of Nuclear Physics (SINP MSU), 1(2), Leninskie gory, GSP-1, 119991 Moscow, Russia}

\author{I.A. Vaiman}
\email[]{gosha.vaiman@gmail.com}
%\thanks{}
\affiliation{Federal State Budget Educational Institution of Higher Education, M.V. Lomonosov Moscow State University, Department of Physics, 1(2), Leninskie gory, GSP-1, 119991 Moscow, Russia}
\affiliation{Federal State Budget Educational Institution of Higher Education, M.V. Lomonosov Moscow State University, Skobeltsyn Institute of Nuclear Physics (SINP MSU), 1(2), Leninskie gory, GSP-1, 119991 Moscow, Russia}


\date{\today}
\begin{abstract}
Primary very high energy $\gamma$-rays from $\gamma$-ray bursts (GRBs) are partially absorbed on extragalactic background light (EBL) photons with subsequent formation of intergalactic electromagnetic cascades. Characteristics of the observable cascade $\gamma$-ray signal are sensitive to the strength and structure of the extragalactic magnetic field (EGMF). GRB~190114C was recently detected with the MAGIC imaging atmospheric Cherenkov telescopes, for the first time allowing to estimate the observable cascade intensity. We inquire whether any constraints on the EGMF strength and structure could be obtained from publicly-available $\gamma$-ray data on GRB~190114C. We present detailed calculations of the observable cascade signal for various EGMF configurations. We show that the sensitivity of the Fermi-LAT space $\gamma$-ray telescope is not sufficient to obtain such constraints on the EGMF parameters. However, next-generation space $\gamma$-ray observatories such as MAST would be able to detect pair echoes from GRBs similar to GRB~190114C for the EGMF strength below $10^{-17}-10^{-18}$ G.
\end{abstract}
%\keywords{}
\maketitle
\section{Introduction}

The recent detection of very high energy (VHE, $E > 100$~GeV) $\gamma$-rays from $\gamma$-ray bursts (GRBs) with imaging atmospheric Cherenkov telescopes (IACTs) MAGIC \cite{MAGIC2019} and H.E.S.S. \cite{Abdalla2019} have aroused great interest. Besides being important for the understanding of ``intrinsic'' physics of GRBs (e.g. \cite{Derishev2019,Ravasio2019,Wang2019,Fraija2019a,Fraija2019b, Wang2019_arxiv}), these observations could in principle be used to constrain the spectrum of the extragalactic background light (EBL) as well as the strength and structure of the extragalactic magnetic field (EGMF). Primary VHE $\gamma$-rays escaping from the source are partially absorbed on EBL photons by means of the pair production (PP) process ($\gamma\gamma \rightarrow e^{+}e^{-}$) \cite{Nikishov1962,Gould1967}. This leads to a characteristic cutoff in the spectra of distant extragalactic sources \cite{Fazio1970}. The imprint of the EBL in the spectra of blazars was robustly detected with the Fermi-LAT space $\gamma$-ray telescope \cite{Ackermann2012, Fermi-LAT2018} and IACTs (e.g. \cite{Abramowski2013}). Secondary electrons and positrons (hereafter ``electrons'' for simplicity) get deflected in the EGMF and then produce cascade $\gamma$-rays by means of the inverse Compton (IC) process $e^{-} \gamma \rightarrow e^{-'} \gamma^{'}$ or $e^{+} \gamma \rightarrow e^{+'} \gamma^{'}$. Parameters of the observable $\gamma$-ray flux are sensitive to the EGMF strength and structure \cite{Plaga1995,Neronov2009,Neronov2010}.

The first lower limits on the EGMF strength ($B\ge 3\cdot10^{-16}$ G) obtained with Fermi-LAT \cite{Atwood2009} data on blazars using spectral information \cite{Neronov2010} were subsequently found to be subject to significant systematic effects including those related to the unknown duty cycle \cite{Dermer2011,Taylor2011} and poorly constrained spectral properties of the source, uncertainties of the EBL spectrum, etc. \cite{Arlen2014,Finke2015}. In particular, \cite{Arlen2014} concluded that it is hard to rule out the zero-EGMF hypothesis. Under such circumstances, an independent channel of information is desirable such as that provided by GRB observations at high ($E > 100$~MeV) and very high energies \cite{Plaga1995,Dai2002,Ichiki2008,Murase2009,Takahashi2010,Veres2017}.

The observable signal from intergalactic electromagnetic cascades depends on many parameters, including the primary intensity, the duration of the flare, the shape of the intrinsic spectrum, and the redshift of the source. In the present paper we perform detailed numerical calculations of the observable intergalactic cascade signal from GRB~190114C taking into account statistical and systematic uncertainties of the intrinsic VHE $\gamma$-ray spectrum (i.e. the spectrum of $\gamma$-rays that have escaped into the intergalactic medium). We also take into account the systematic uncertainty of the EBL intensity, which appears to be a major factor affecting the observable pair echo intensity. The inclusion of any other systematic effect is equivalent to the addition of a nuisance parameter, further increasing the uncertainty of the EGMF parameter measurement, and thus reinforcing our conclusions.

This paper is organized as follows. In Sect.~\ref{sect:qualitative} we describe the approach to constraining the EGMF parameters with pair echos from GRBs, in Sect.~\ref{sect:spectrum} we reconstruct the intrinsic $\gamma$-ray spectrum of GRB~190114C (the optimization procedure presented in this section is based on \cite{Dzhatdoev2017a}). In Sect.~\ref{sect:fermi} we describe our analysis of Fermi-LAT data in order to set experimental upper limits on the pair echo intensity from GRB~190114C. In Sect.~\ref{sect:cascade} we present our results for the expected observable intergalactic cascade signal assuming various values of $B$. We conclude that the sensitivity of the Fermi-LAT telescope is not sufficient to constrain the EGMF.

Very recently, Wang et al. \cite{Wang2020} (hereafter W20) found that observations of GRB~190114C with MAGIC and Fermi-LAT allow one to rule out $B<3\cdot10^{-20}$~G. These results were derived for a large-scale EGMF with the coherence length $L_{c}$ much greater than the characteristic cascade electron energy loss length $L_{E-e}$. Our conclusions are significantly different from those of W20 (see Sect.~\ref{sect:discussion} for a more detailed discussion). Finally, we conclude in Sect.~\ref{sect:conclusions}.

\section{Qualitative considerations and analytic estimates \label{sect:qualitative}}

Above the observable energy of $\approx200$ GeV, \mbox{$\gamma$-rays} from GRB~190114C (redshift $z = 0.4245$) are strongly absorbed on the EBL, i.e. the optical depth of the PP process becomes greater than unity. The secondary electrons are produced with the typical angular spread $\theta_{P} \approx 1/\gamma_{e}$, where $\gamma_{e} = E_e / (m_e c^2)$ is the electron Lorentz factor, $E_e$ is the electron energy, $m_e$ is the electron mass, $c$ is the speed of light. The electrons propagate through the EGMF, get deflected by the angle $\theta_{B}$ and accumulate time delay, meanwhile producing cascade $\gamma$-rays via the IC acts.

For the range of parameters considered in the present work, pair production occurs mainly on EBL photons, while IC scattering --- mainly on CMB photons \cite{Aharonian2002}. The contribution of IC scattering on EBL photons to the observable cascade spectrum is significant only in the cutoff region of this spectrum (see \cite{Dzhatdoev2017c}, Fig. 3). A very short and basic introduction to the physics of intergalactic electromagnetic cascades is available in \cite{Dzhatdoev2017b}; much more detailed treatments could be found in \cite{Neronov2009,Berezinsky2016}. The geometry of intergalactic electromagnetic cascades in the magnetized Universe was also considered in e.g. \cite{Dolag2009,Fitoussi2017}.

Following \cite{Neronov2009} (see their eq. (39)), the time delay of cascade $\gamma$-rays for the typical conditions considered in the present work may be estimated as:
\begin{equation}
\Delta T= \sqrt{\Delta T_{B}^{2}+\Delta T_{P}^{2}},
\end{equation}

\begin{equation}
\Delta T_{B}= \frac{(1+z)L_{\gamma}}{2c}\left( 1- \frac{L_{\gamma}}{L_{s}}\right) \theta_{B}^{2},
\end{equation}

\begin{equation}
\Delta T_{P}= \frac{(1+z)L_{\gamma}}{2c}\left( 1- \frac{L_{\gamma}}{L_{s}}\right) \theta_{P}^{2},
\end{equation}
where $L_{\gamma}$ is the distance travelled by the parent $\gamma$-ray before interaction and $L_{s}$ is the distance from the source to the observer. According to \cite{Neronov2009} (eq. (30)) for a large-scale EGMF and $L_{\gamma} \ll L_{s}$:
\begin{equation}
\theta_{B} \approx \frac{L_{E-e}}{R_{L}} \approx \frac{3\cdot10^{-6}}{(1+z)^{2}} \left(\frac{B}{10^{-18} G}\right) {\left(\frac{E_{e}}{10 TeV}\right)}^{-2},
\end{equation}
where $L_{E-e}$ is the characteristic cascade electron energy loss length, $R_{L}$ is the electron Larmor radius, $z$ is the redshift. For the conditions of our work ($B>10^{-20}$ G and $E_{\gamma}<10$ GeV, where Fermi-LAT has the maximal sensitivity), as a rule, $\theta_{B}>\theta_{P}$ and thus $\Delta T_{B}>\Delta T_{P}$ (see \cite{Neronov2009}, eq. (41)). Therefore, in this case $\Delta T \sim \Delta T_{B} \propto B^{2}$. If the time delay exceeds the observation time of the Fermi-LAT instrument ($\Delta T > \Delta T_{\mathrm{obs-LAT}}$), the observable cascade spectrum will exhibit a cutoff at low energies, since low energy cascade $\gamma$-rays acquire a greater time delay. Given that the typical energy of cascade $\gamma$-rays $E_{\gamma-c} \propto E_{e}^{2}$ \cite{Neronov2009}, the energy of the cutoff $E_{br} \propto B$.

\begin{figure}
\vspace{0.2cm}
\includegraphics[width=9cm]{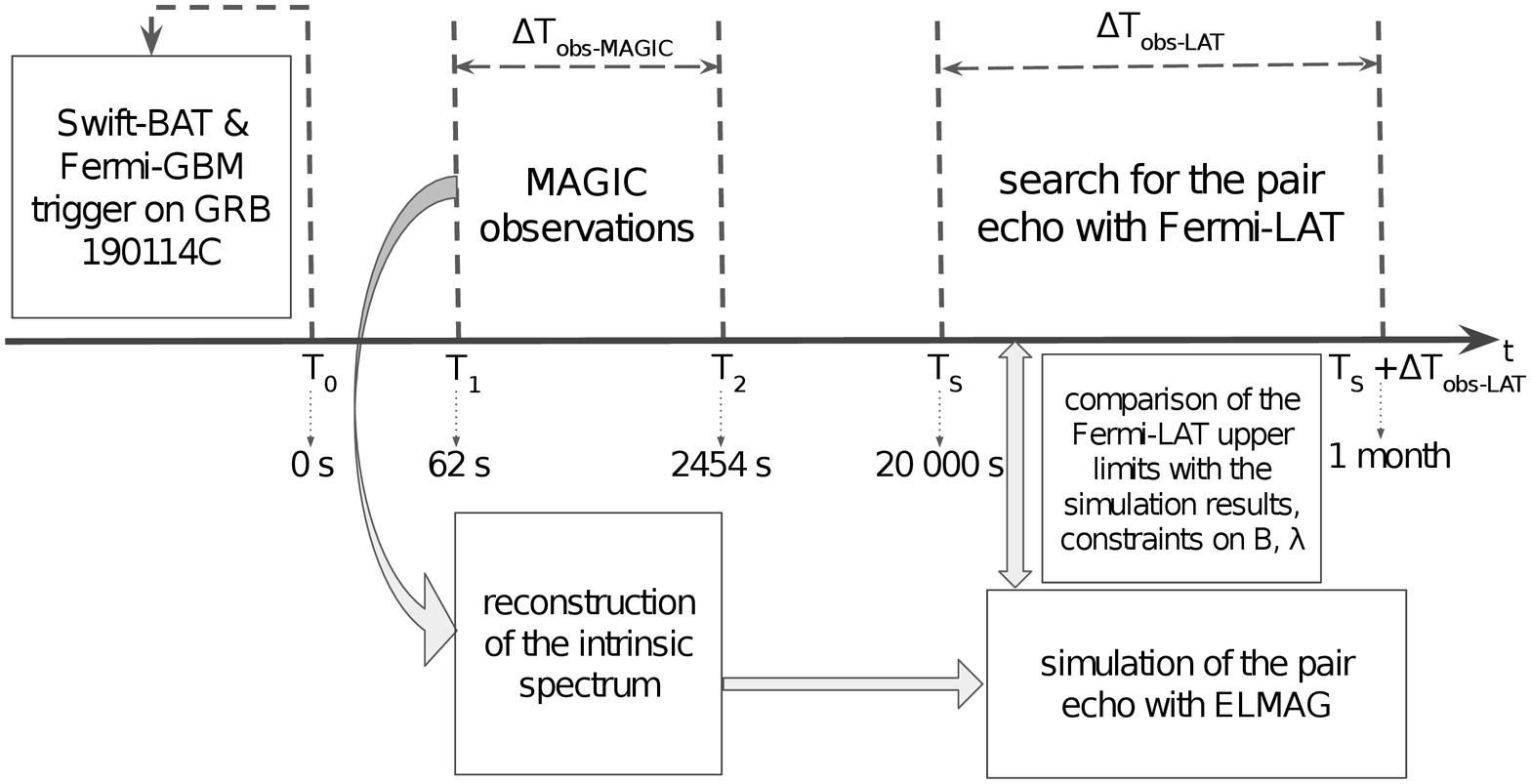}
\caption{A flowchart of the present work. \label{fig:time_scales}}
\end{figure}

Our approach to simulation of the pair echo from GRB 190114C is illustrated in Fig. \ref{fig:time_scales}.The observable VHE $\gamma$-ray intensity $dN_{e}/dE$ (the subscript $e$ stands for ``experimental'') of GRB~190114C in the energy range of 0.2--1.1 TeV was measured with the MAGIC IACTs over the time period between $T_{0}+T_{1}$ and $T_{0}+T_{2}$ \cite{MAGIC2019}, where $T_{0}$ is the trigger time provided by the Burst Alert Telescope (BAT) onboard the Neil Gehrels Swift Observatory \cite{Gropp2019} and the $\gamma$-ray Burst Monitor (GBM) onboard the Fermi satellite \cite{Hamburg2019}, $T_{1} = 62$~s, and $T_{2} = 2454$~s. The corresponding spectral energy distribution (SED~=~$E^{2}dN_{e}/dE$) is shown in Fig.~\ref{fig:MAGIC} together with its statistical uncertainties as red circles with error bars. We use the MAGIC observations of GRB~190114C to reconstruct the intrinsic spectrum of the source (see Sect. \ref{sect:spectrum}). We search for the pair echo with Fermi-LAT starting at $T_0 + T_S$, where $T_S = 2\cdot10^{4}$~s, and derive upper limits on the pair echo SED (Sect. \ref{sect:fermi}). The time shift $T_S$ is needed in order to avoid the contamination from the afterglow $\gamma$-rays. The reconstructed intrinsic VHE spectrum is used as an input spectrum in our simulation of the pair echo from GRB~190114C (Sect. \ref{sect:cascade}), performed with the publicly-available code ELMAG 3 \cite{Blytt2020, Kachelriess2012}. Finally, we compare the resulting model intensity of the observable cascade SED with the Fermi-LAT upper limits.

\section{Reconstruction of the intrinsic spectrum \label{sect:spectrum}}

In what follows we characterize the intrinsic VHE $\gamma$-ray intensity with a simple form $\propto E^{-\gamma}exp(-E/E_{c})$. This is a reasonable assumption given the narrow energy range of the MAGIC spectrum (less than one order of magnitude). Assuming the EBL model of Gilmore at al. \cite{Gilmore2012} (hereafter G12) and the redshift of the source $z = 0.4245$ \cite{Selsing2019, Castro-Tirado2019}, we estimate the spectral parameters $\gamma$ and $E_{c}$ as follows.

We set ($\gamma = 2$, $E_{c} = 10$~TeV) as the initial values of these parameters and calculate the observable intensity in all four bins of the MAGIC spectrum accounting for the effect of intergalactic absorption, dividing each of these bins to $N_{div} = 21$ parts in order to ensure small variation of the intergalactic $\gamma\gamma$ optical depth $\tau$ over any of these 84 new narrow bins.
We define the $\chi^{2}$ functional form as follows:
\begin{equation}
\chi^{2}(\gamma,E_{c}) = \sum_{i=1}^{N_{bin}}{\frac{\left(F_{m;i}(\gamma,E_{c})-F_{e;i}\right)^{2}}{\Delta_{i}^{2}}},
\end{equation}
where $N_{bin} = 4$ is the number of MAGIC energy bins, $F_{e;i}$ and $F_{m;i}$ are the measured (experimental) and model SEDs, respectively, normalized to their values at the decorrelation energy $E_{d}$:
\begin{equation}
F_{e;i} = \frac{E_{i}^{2}(dN_{e}/dE)_{i}}{E_{d}^{2}(dN_{e}/dE)_{d}},
\end{equation}
\begin{equation}
F_{m;i}(\gamma,E_{c}) = \frac{F_{m}(E_{min;i},E_{max;i},\gamma,E_{c})}{F_{m}(E_{min;d},E_{max;d},\gamma,E_{c})},
\end{equation}
where
\begin{equation}
F_{m}(E_{min;i},E_{max;i},\gamma,E_{c})= \frac{\sum_{j=0}^{N_{div}}{E_{j}^{2-\gamma}e^{-E_{j}/E_{c}}e^{-\tau(E_{j},z)}}}{N_{div}},
\end{equation}
and
\begin{equation}
E_{j} = E_{min;i}+ (E_{max;i}-E_{min;i})\frac{j}{N_{div}}.
\end{equation}

$E_{min;i}$ and $E_{max;i}$ are the minimal and maximal energies of every MAGIC energy bin, respectively; decorrelation energy is defined as the central energy of the bin with the minimal statistical uncertainty of the measured spectrum; the minimal and maximal energies of this bin are denoted as $E_{min;d}$ and $E_{max;d}$, respectively. Finally, $\Delta_{i}$ is the statistical uncertainty of the measured SED in the $i^{th}$ energy bin; it was normalized in the same way as the measured SED.

\begin{figure}
\vspace{0.2cm}
\includegraphics[width=9cm]{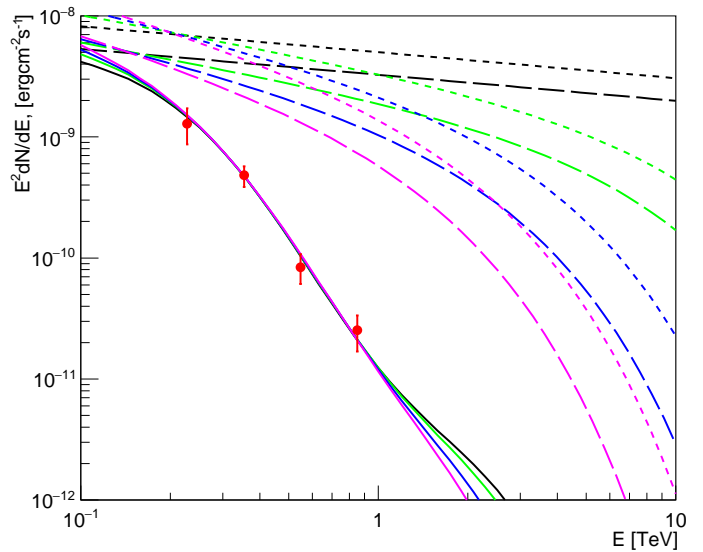}
\caption{SED of GRB~190114C measured with MAGIC (red circles with statistical uncertainties) together with best fits for various EBL model options (solid curves), the corresponding intrinsic SEDs (long-dashed curves), and the blueshifted intrinsic SEDs (short-dashed curves); the SEDs are averaged over the time period from $T_{1}$ to $T_{2}$, see text for more details. \label{fig:MAGIC}}
\end{figure}

Then, varying the fitting parameters ($\gamma$,$E_{c}$) and repeating the above-described procedure for every new set of these parameters, we minimize the $\chi^{2}$ form with the MINUIT package \cite{James1975} integrated into the ROOT framework and determine the best-fit values of $\gamma$ and $E_{c}$. The corresponding best-fit observable SED is shown in Fig.~\ref{fig:MAGIC} as black solid curve.

The intrinsic VHE $\gamma$-ray spectrum resulting from this procedure is also shown in Fig.~\ref{fig:MAGIC} as black long-dashed curve. Finally, we account for the effect of redshift \footnote{here we ensure that the number of primary $\gamma$-rays is conserved}; the resulting intrinsic VHE $\gamma$-ray spectrum in the source rest frame is shown in Fig.~\ref{fig:MAGIC} as black short-dashed curve. We note that for the G12 EBL model the best-fit value $E_c \rightarrow +\infty$, i.e. the intrinsic spectrum does not reveal a cutoff. On the other hand, the most natural models of GRB emission predict the existence of such a cutoff due to the PP process inside the source (see e.g. \cite{Veres2019}). This apparent slight tension between the reconstructed and predicted intrinsic spectral shapes could be relaxed if we assume a model of EBL with a slightly diminished intensity compared to the ``nominal'' G12 EBL model.

We repeat the whole procedure of the intrinsic spectrum reconstruction for three different normalizations of the EBL intensity, namely, those of 90 \%, 80 \%, and 70 \% of the original EBL intensity according to the G12 EBL model. The results for these three runs of the optimization procedure are shown in Fig.~\ref{fig:MAGIC} as green, blue, and magenta curves, respectively.

\section{Fermi-LAT data analysis \label{sect:fermi}}

Here we derive upper limits on the observable SED of GRB~190114C. We select Fermi LAT data within 1 month of observation, starting at $T_{0}+T_{S}$, where $T_{S}= 2\cdot10^{4}$~$\mathrm{s}$. The region of interest (ROI) is a circle with the radius of $12^{\circ}$, centered at the position of the GRB ($\alpha_{\mathrm{J2000}} = 54.507^{\circ}, \delta_{\mathrm{J2000}} = -26.947^{\circ}$). We have applied the energy selection from $100~\mathrm{MeV}$ to $100~\mathrm{GeV}$. For other selection parameters we use standard recommendations for off-plane point source identification with Fermi-LAT.

We then perform unbinned likelihood analysis of the selected data with Fermitools \cite{FSSC}. We construct a model of observed emission including the following sources that could contribute to the detected $\gamma$-ray counts inside the ROI: 1) GRB~190114C itself, modelled as a point-like source with power-law spectrum at the center of the ROI, 2) all sources from Fermi 8-Year Point Source Catalog (4FGL) \cite{Abdollahi2020} located within $17^{\circ}$ from the center of the ROI, 3) galactic and isotropic diffuse $\gamma$-ray backgrounds using models provided by the Fermi-LAT Collaboration. For GRB~190114C we set both spectral index and normalization as free parameters; for point-like sources within $5^{\circ}$ from the center of the ROI and the diffuse backgrounds only the normalizations were left free, while the spectral shapes were fixed; for point-like sources beyond $5^{\circ}$ from the center of the ROI both normalizations and shapes were fixed.

Using this model of the observed emission, we perform the maximization of the likelihood, i.e. we determine the values of parameters which yield the maximal probability of producing the observed $\gamma$-ray counts. We calculate the value of the test statistic corresponding to the hypothesis of the GRB~190114C emission being present in the dataset against the null hypothesis of it being absent. The resulting value of the test statistic $TS \ll 1$, which means that there is no significant $\gamma$-ray flux detected from this GRB.

Given that no signal from GRB~190114C was detected we place upper limits on its SED. At this stage of our analysis we reduce the emission model to only four sources: GRB~190114C, 4FGL J0348.5-2749 (the brightest point-like source inside the ROI), and two diffuse backgrounds described above. These sources are responsible for almost all observed \mbox{$\gamma$-rays} inside the ROI. Finally, we run the user-contributed Python script likeSED.py \cite{JohnsonFSSC} to calculate upper limits (95\% C.L.) on the emission from GRB~190114C in six energy bins (two bins per decade of energy). These upper limits are shown in Fig.~\ref{fig:constraints100}. They are slightly different from similar results of W20.
\vspace{0.5cm}

\section{Simulation of pair echo \\ from GRB~190114C \label{sect:cascade}}

Using the publicly-available code ELMAG 3 \cite{Blytt2020,Kachelriess2012}, we calculate the observable SED of intergalactic cascades over $\Delta T_{\mathrm{obs-LAT}} = 1$ month assuming the ``original'' G12 EBL model (the corresponding intrinsic VHE $\gamma$-ray SED over the time period $\Delta T_{\mathrm{obs-MAGIC}} = T_{2}-T_{1}$ is shown as short-dashed black line in Fig.~\ref{fig:MAGIC}) The EGMF was modelled following the approach of \cite{Giacalone1994,Giacalone1999} (see Subsect. 2.1 of \cite{Blytt2020}) as isotropic random non-helical turbulent field with a Kolmogorov spectrum and Gaussian variance $B_{RMS}$ (hereafter simply $B$). In total, 200 field modes were simulated with the minimal and maximal spatial scales $L_{min} = 5\cdot10^{-4}$~Mpc and $L_{max} = 5$~Mpc, respectively. In this case, the coherence length $L_{c}$= 1 Mpc \cite{Blytt2020}. The cascade electrons were propagated through the EGMF as described in Subsect. 2.2 of \cite{Blytt2020} (i.e. using the full three-dimensional simulation).

For simplicity, we chose to calculate models of pair echo emission over the time period from $T_{0}$ to $T_{0} + 1$ month rather than starting at $T_{0} + T_{S}$. Subtracting $\gamma$-rays that have time delay less than $T_{S}$ would decrease the observable intensity and thus would reinforce our conclusions. The ELMAG 3 code includes two terms of time delay arising from the deflection of cascade electrons in the EGMF and from the angular spread of electrons in PP acts, but it does not account for another two terms, namely, the one arising from the angular spread of cascade $\gamma$-rays in IC acts, as well as the accumulated deflection of cascade electrons in IC acts (``cascade electron recoil''). A more detailed account of these effects is underway and will be published elsewhere. Given the difference of $\Delta T_{\mathrm{obs-MAGIC}}$ and $\Delta T_{\mathrm{obs-LAT}}$, an additional factor $K_{Corr}= \Delta T_{\mathrm{obs-MAGIC}}/\Delta T_{\mathrm{obs-LAT}}$ was introduced in order to obtain the observable SED in the Fermi-LAT energy band.

\begin{figure}
\vspace{0.2cm}
\includegraphics[width=9cm]{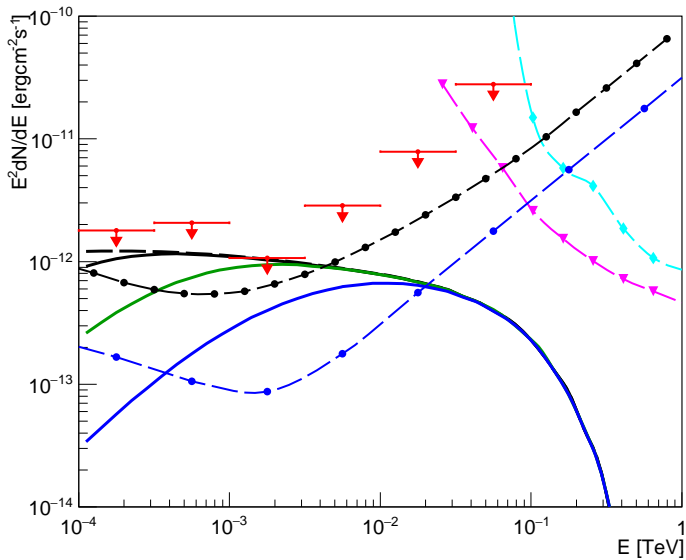}
\caption{Upper limits on the SED of GRB~190114C derived from Fermi-LAT data (red horizontal bars with downwards arrows) together with model SEDs for various values of $B$ (curves without symbols; see text for more details). The SEDs are averaged over the Fermi-LAT observation time $\Delta T_{\mathrm{obs-LAT}}$. Also shown are the differential sensitivity of the CTA IACT array and the MAST $\gamma$-ray telescope (filled symbols connected with long-dashed lines; see text for more details). \label{fig:constraints100}}
\end{figure}

The resulting observable cascade SEDs are shown in Fig.~\ref{fig:constraints100} for four different values of $B = 10^{-20}$~G (black solid curve), $10^{-19}$~G (green solid curve), $10^{-18}$~G (blue solid curve), and $B = 0$ (black dashed curve). All these curves are below the Fermi-LAT upper limits. Therefore, no constrains on the EGMF strength and/or structure could be set using these data. The account of the MAGIC systematics on the intrinsic spectrum normalization (about 50 \%) would introduce an additional source of uncertainty \cite{MAGIC2019}. We note that all four model curves practically coincide at $E > 40$~GeV, while at lower energies three of these curves successively branch down from the zero-field curve at an energy $E_{br} \approx 40$~GeV$\cdot B/(10^{-18}  \mathrm{\,G})$. This behaviour of observable SEDs is in full agreement with analytic estimates presented in Sect.~\ref{sect:qualitative}.

The sensitivity of the CTA IACT array \cite{Acharya2013,CTA2018} for for five hours of observation, the statistical significance $Z= 5 \sigma$, and five energy bins per decade of energy is also shown in Fig.~\ref{fig:constraints100} for the zenith angle of 20$^{\circ}$ (magenta triangles connected with magenta dashed curve) and of 60$^{\circ}$ (cyan diamonds connected with cyan dashed curve). The energy threshold of CTA is too high to detect the cascade signal.

Finally, in Fig.~\ref{fig:constraints100} we present the sensitivity of the MAST projected space $\gamma$-ray telescope \cite{Dzhatdoev2019a} for one month of observation in the survey mode and two following options: 1) five energy bins per decade of energy and $Z= 5 \sigma$ (black circles connected with black dashed curve), 2) two energy bins per decade of energy and $Z= 2 \sigma$ (blue circles connected with blue dashed curve). For both options we have imposed an additional condition that the expected number of counts from the pair echo in every energy bin is greater than unity. We note that observations with MAST would allow to probe the EGMF $B<10^{-17}$~G using the pair echo method. Stronger values of $B = 10^{-14}-10^{-17}$~G could be probed if a ``magnetically broadened cascade pattern'' \cite{Abramowski2014} could be detected \cite{Neronov2009,Dzhatdoev2018,Dzhatdoev2019a}.

\begin{figure}
\vspace{0.2cm}
\includegraphics[width=9cm]{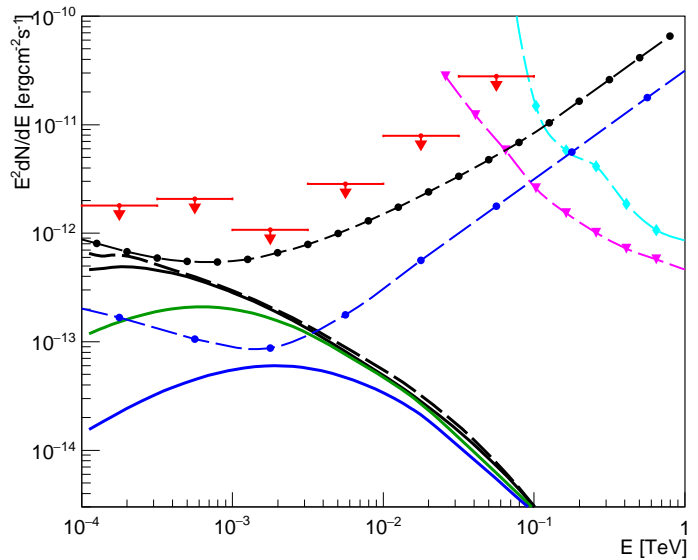}
\caption{Same as in Fig.~\ref{fig:constraints100}, but model curves are for the modified EBL with $K_{EBL} = 0.7$. \label{fig:constraints70}}
\end{figure}

We have also performed similar calculations for a modified EBL model with the normalization factor $K_{EBL} = 0.7$ (see Fig.~\ref{fig:constraints70}). The corresponding intrinsic VHE $\gamma$-ray SED over the time period $\Delta T_{\mathrm{obs-MAGIC}}= T_{2}-T_{1}$ for this EBL model is shown as short-dashed magenta line in Fig.~\ref{fig:MAGIC}. The residual difference between four model curves in Fig.~\ref{fig:constraints70} at high energies is mainly due to statistical fluctuations. Fig.~\ref{fig:constraints70} demonstrates that for $K_{EBL} = 0.7$ the model pair echo intensity is well below the Fermi-LAT upper limits even for $B = 0$. However, observations with MAST would still allow to probe the range of $B < 10^{-18}$ G using the pair echo method.

Qualitatively similar results to those presented in Figs.~\ref{fig:constraints100}--\ref{fig:constraints70} could be obtained with the publicly-available code of \cite{Fitoussi2017}. We note that our results apply directly to a large-scale EGMF. For a small-scale EGMF with $L_{c}<L_{E-e}$ a stronger magnetic field is required in order to achieve the same deflection of cascade electrons: $B\propto \sqrt{L_{E-e}/L_{c}}$ (e.g. \cite{Neronov2009}). The dependence of $L_{E-e}$ on the electron energy $E_{e}$ for $z=0$ was presented in \cite{Dzhatdoev2019b} in Fig. 2 (left); this figure was produced assuming the approximation for the IC process obtained in \cite{Khangulyan2014}. At $E_{e}<10$ TeV the Thomson approximation of \cite{Neronov2009} is applicable (see their eq. (28)). At $z=0$ $L_{E-e}\approx3$ Mpc for $E_{e}=100$ GeV and $L_{E-e}\approx80$ kpc for $E_{e}=5$ TeV; $L_{E-e}(E_{e},z)/L_{E-e}(E_{e},0)\propto(1+z)^{-4}$.

In the present work we have neglected the emission from primary afterglow after $T_{S}$. Preliminary estimates show that the inclusion of this emission does not change our conclusions. A more detailed account of the primary afterglow is underway and will be published elsewhere. We note that our results are not sensitive to the time distribution of very high energy $\gamma$-rays inside the MAGIC time window, because this time window is much narrower than the Fermi-LAT time window.

\section{Discussion \label{sect:discussion}}

\subsection{Influence of the EBL normalization \\ on the pair echo intensity}

The typical energy of cascade $\gamma$-rays $E_{\gamma-c}\approx (4/3)\gamma_{e}^{2}\epsilon$, where $\epsilon\approx 6.3\cdot10^{-4}$ eV is the characteristic energy of background photons \cite{Berezinsky2016}\footnote{Under the conditions of the present work cascade $\gamma$-rays are produced mostly on cosmic microwave background photons and intergalactic cascades have only one dominant generation}. Under the conditions of our analysis Fermi-LAT is most sensitive to the pair echo emission at $E \approx 1$ GeV corresponding to the primary $\gamma$-ray energy $E_{\gamma}\approx 1.2$ TeV. The primary $\gamma$-ray intensity at such energies is about five times greater for $K_{EBL} = 1$ than for $K_{EBL} = 0.7$, explaining a decrease in intensity of cascade $\gamma$-rays at $E = 1$ GeV by a comparable factor \footnote{Additionally, the fraction of survived $\gamma$-rays for $K_{EBL} = 0.7$ is somewhat greater than for $K_{EBL} = 1$, again leading to a decrease of the total energy transferred to cascade electrons}.

Thus, a modest (30 \%) change of the EBL normalization corresponds to a strong (an order of magnitude) decrease of the observable intensity in the energy region where Fermi-LAT has the maximal sensitivity (1--3 GeV under the conditions of the present work). Dedicated studies of the EBL indeed suggest that the total uncertainty of $K_{EBL}$ may be around 30 \% \cite{Korochkin2018} or even greater \cite{Stecker2016}. Theoretical  models of the EBL \cite{Kneiske2004,Primack2005,Franceschini2008,Kneiske2010,Dominguez2011}, G12, \cite{Inoue2013,Franceschini2017,Franceschini2018} also reveal a significant spread of the predicted intensity amounting to dozens of percent (see Fig. 7 in \cite{Inoue2013}).

\subsection{Possible influence of plasma energy losses \\ and other effects on the pair echo intensity}

Pair beams resulting from the development of intergalactic electromagnetic cascades may be subject to plasma instabilities that may cause additional energy losses \cite{Broderick2012}. At present it is unclear whether these ``plasma losses'' are considerable or subdominant with respect to inverse Compton (IC) losses (e.g. \cite{Schlickeiser2012,Miniati2013,Chang2014,Sironi2014,Menzler2015,Kempf2016,Vafin2018,Vafin2019}). Therefore, in the present work we have accounted for only the IC losses. We note, however, that the inclusion of the plasma losses would decrease the pair echo intensity and thus would reinforce our conclusions. Finally, we note that the inclusion of any additional effect such as $\gamma \rightarrow ALP$ oscillations, Lorentz invariance violation, or an account for the possibility that a part of VHE $\gamma$-rays observed with MAGIC are in fact not primary, but cascade ones, is tantamount to the inclusion of a new nuisance parameter, further increasing the uncertainty of the EGMF parameter measurement, and thus reinforcing our conclusions.

	\subsection{Comparison with W20}

The present study has a number of differences with respect to W20 in both model assumptions and calculation techniques. From the text of W20 it transpires that they did not perform a detailed reconstruction of the intrinsic $\gamma$-ray spectrum in the TeV energy range, as was done in the present work (see our Sect.~\ref{sect:spectrum}). Instead, W20 assumed that this primary spectrum had the power-law index $\gamma=-2$, and that a power-law decay of intensity starts at 6 s. The latter assumption is not supported with the model developed in \cite{Veres2019} (see their Extended Data Fig. 7) where the maximum at the 300 GeV -- 1 TeV light curve (dark green curve) is situated at $\approx$20~s and not at 6~s. Additionally, the observable intensity of cascade $\gamma$-rays significantly depends on the shape of the primary spectrum even if the total energy of primary $\gamma$-rays is fixed.

Furthermore, we found that according to this model the total energy radiated in the afterglow phase of GRB~190114C in the 300 GeV -- 1 TeV energy range exceeds the energy output in the same energy range inside the time window from $T_{0}+T_{1}$ to $T_{0}+T_{2}$ by the factor of only $K_{T} = 2.4$ and not by $K_{T} = 5$ as was claimed in W20 \footnote{the internal opacity at E= 1 TeV could be significantly larger than at 300 GeV; an account of this effect would further decrease the value of $K_{T}$}. Therefore, we argue that W20 have significantly overestimated the normalization of the observable pair echo intensity. They also did not account the EBL uncertainty that could decrease the observable pair echo flux at $E = 1$ GeV by the factor of five (see discussion above). Some additional very high energy $\gamma$-rays could in principle come from the prompt emission phase of GRB 190114C. However, the internal opacity for these prompt $\gamma$-rays is expected to be high. Indeed, the authors of \cite{Veres2019} demonstrate that the optical depth for TeV $\gamma$-rays is significant even in the time interval 68--110 s after the trigger time (see their Fig. 3). Much stronger absorption is expected for the prompt emission phase $\gamma$-rays. For this reason we did not include these prompt $\gamma$-rays in our calculations.

Finally, we note that W20 did not calculate arrival time for individual observable $\gamma$-rays, but introduced a normalization factor $t_{dur}$ in order to compute the observable flux (see their eq. (4)). This procedure is not directly comparable to Monte Carlo approach utilized in the present paper.

\section{Conclusions \label{sect:conclusions}}

The sensitivity of the Fermi-LAT $\gamma$-ray telescope is not sufficient to detect the intergalactic electromagnetic cascade signal from GRB~190114C over the time period of one month. The calculations for different values of $\Delta T_{\mathrm{obs-LAT}}$ are straightforward; the results of these caclulations will be reported elsewhere. CTA will not be able to detect pair echo from GRBs similar to GRB~19014C due to a relatively high energy threshold of this $\gamma$-ray detector compared to space $\gamma$-ray telescopes. However, observations with CTA would be crucial for constraining the shape of the intrinsic spectrum. Hopefully, future $\gamma$-ray detectors such as MAST \cite{Dzhatdoev2019a} with much improved sensitivity will be able to probe the EGMF strength and structure for $B < 10^{-17}-10^{-18}$ G using the pair echo method.

\begin{acknowledgments}
We acknowledge helpful discussions with Prof. K.~Murase and Dr. Z.-R.~Wang. This work was supported by the Russian Science Foundation (RSF) (project no. 18-72-00083). This research has made use of publicly-available Fermi-LAT experiment data, the CTA instrument response functions provided by the CTA Consortium and Observatory (version prod3b-v2, see \url{http://www.cta-observatory.org/science/cta-performance/} for more details), and the NASA ADS and INSPIRE bibliographical systems. All graphs in the present paper were produced with the ROOT software toolkit \cite{Brun1997}. We are grateful to the organizers of the researcher school ``Multimessenger data analysis in the era of CTA'' (Sexten, Italy, 2019) for tutorials provided by them. E.~I.~Podlesnyi thanks the Foundation for the Advancement of Theoretical Physics and Mathematics ``BASIS'' for the support in participation at the aforementioned school (travel-grant no. 19-28-030) and for the student scholarship (agreement no. 19-2-6-195-1).
\end{acknowledgments}

\bibliography{GRB-EGMF}
\end{document}